\documentclass[twocolumn,secnumarabic,amssymb, nobibnotes, aps, prr]{revtex4-2}
\usepackage{braket}
\usepackage{physics}
\usepackage{mathtools}
\usepackage{amsfonts}
\usepackage{graphicx}
\usepackage{dcolumn}
\usepackage{bm}
\usepackage{diagbox}

\usepackage{float}
\usepackage{xcolor}
\usepackage{soul} 
\usepackage{placeins} 

\newsavebox{\mstrut}

\newcommand{\scom}[1]{%
    \sbox{\mstrut}{\(#1\)}%
    \mathinner{\left[\mkern-2mu\left[{#1}\right]\kern-0.1\ht\mstrut\right]}%
}

\newcommand{\up}{|\!\!\uparrow\rangle}
\newcommand{\down}{|\!\!\downarrow\rangle}
\begin{document}

\title{Discrete-time quantum walks in synthetic dimensions}%

\author{Piergiorgio Ferraro}
\email[]{piergiorgio.ferraro@edu.unito.it}
\affiliation{Department of Physics, University of Turin, via P. Giuria 1, 10125 Turin, Italy}
\author{Caio B. Naves}%
\email[]{caio.naves@fysik.su.se}
\affiliation{Department of Physics,
Stockholm University, AlbaNova University Center, 106 91 Stockholm,
Sweden}
\author{Jonas Larson}%
\email[]{jolarson@fysik.su.se}
\affiliation{Department of Physics, Stockholm University, AlbaNova University Center, 106 91 Stockholm, Sweden}
\begin{abstract}
In this work we introduce discrete-time quantum walks in state space, more precisely on \textit{Fock-state lattices}. Fock-state lattices provide a natural and clean setting for implementing lattice models, particularly in quantum optical systems. Thus, contrary to the common setting where the walker resides in real space or phase space, here the walk takes place in a synthetic space. We present a general formalism based on Lie algebras and their properties. For each Lie algebra one can associate both a phase space and a Fock-state lattice, and by understanding how these spaces are related, together with the action of generalized displacement operators, we construct the discrete unitary operator that generates the walk. In this framework the displacement operators replace the usual nearest-neighbor shifts and lead to state-dependent tunneling on the lattice. By considering several examples we demonstrate ballistic spreading and other characteristic features of discrete-time quantum walks, such as coin–walker entanglement and symmetry-induced interference patterns. We also show that different algebraic structures can give rise to qualitatively different dynamics, including anomalous behavior such as super-ballistic spreading as well as localization effects.
\end{abstract}

\maketitle


\section{Introduction}

Since their introduction~\cite{aharonov1993quantum}, quantum walks have become an important framework for studying quantum transport and information processing~\cite{kempe2003quantum,venegas2012quantum}. In contrast to classical random walks, which exhibit diffusive spreading, quantum walks display ballistic propagation and therefore explore their underlying space much faster. This property motivated early proposals to use quantum walks to accelerate search algorithms~\cite{shenvi2003quantum,childs2003exponential,childs2004spatial}, and subsequently stimulated broad interest in their use for quantum information tasks. Examples include proposals for universal quantum computation~\cite{childs2009universal,lovett2010universal,qiang2024quantum} and quantum simulation~\cite{plenio2008dephasing}. Beyond algorithmic applications, quantum walks have also proven useful for studying properties of lattices and graphs~\cite{wang2013physical}, including topology and edge states~\cite{kitagawa2010exploring,flurin2017observing,wang2018detecting}, transport in complex networks~\cite{plenio2008dephasing,mohseni2008environment,engel2007evidence}, graph classification~\cite{berry2011two,douglas2008classical}, and quantum sensing~\cite{chen2017quantum,annabestani2022multiparameter}.

Quantum walks are commonly divided into two main classes: continuous-time and discrete-time quantum walks (DTQWs)~\cite{childs2010relationship}. In a DTQW the evolution proceeds stroboscopically through alternating applications of a coin operation and a conditional shift of the walker. DTQWs have been realized experimentally in a variety of physical systems, including trapped ions~\cite{perets2008realization,schmitz2009quantum,zahringer2010realization,huerta2020quantum}, photonic and optical platforms~\cite{peruzzo2010quantum,jeong2013experimental,poulios2014quantum,biggerstaff2016enhancing,wang2019simulating,tang2018experimental,sengupta2025experimental}, cold atoms~\cite{karski2009quantum,xie2020topological,dadras2019experimental}, and circuit QED architectures~\cite{flurin2017observing}. In many implementations the walker is associated with motion in real space, and the conditional translation required for the walk must be engineered through carefully tailored interactions. This can make practical realizations experimentally demanding. For this reason, alternative formulations have been explored in which the walker evolves in momentum or phase space instead of real space~\cite{sanders2003quantum,xue2008quantum,xue2009quantum,schmitz2009quantum,zahringer2010realization,hardal2013discrete,flurin2017observing,dadras2019experimental,duan2022quantum,duan2023quantum}, which can significantly simplify experimental implementations.

In the present work we take a different perspective and consider DTQWs that occur directly in Hilbert space. Writing states in the Fock basis naturally defines a lattice structure of discrete occupation states, which we refer to as a \textit{Fock-state lattice} (FSL). Such lattices arise naturally in a variety of quantum optical systems, including trapped ions and cavity or circuit QED platforms~\cite{larson2021jaynes}. In these systems the individual Fock states and their couplings can often be addressed with high precision, making FSLs experimentally accessible and controllable~\cite{wang2016mesoscopic,cai2021topological,saugmann2023fock,larson2021jaynes,mumford2022meissner,mumford2024gauge,yuan2024quantum,peng2025ideal,sriram2025quantized,zhao2025dark}. In recent years such synthetic lattices have attracted considerable attention, with experiments demonstrating phenomena such as topological effects~\cite{deng2022observing,wu2023observation,wang2024realization,wang2024realizing}, flat-band physics~\cite{yang2024realization}, and synthetic gauge effects including the Aharonov--Bohm effect~\cite{zhang2025synthetic}.

A key advantage of this setting is that the lattice structure emerges directly from the Hilbert space itself, allowing the walker to explore synthetic dimensions that are not limited by physical spatial geometry. In particular, FSLs can naturally realize higher-dimensional structures, enabling DTQWs in dimensions $D>3$, analogous to earlier studies of quantum walks on general graphs~\cite{aharonov2001quantum,moore2002quantum,kendon2006quantum,schuld2014quantum}. The main challenge is then to construct suitable walker operations that generate controlled transitions between Fock states. In this work we address this problem by employing generalized phase-space displacement operators. Phase space thus serves as an implementation framework: the displacement operators generate the step operations, while the walk itself is defined on the discrete FSL.

To develop this approach systematically we formulate the construction of FSL DTQWs using the language of Lie algebras~\cite{georgi2000lie,fuchs2003symmetries,sattinger2013lie}. Each Lie algebra defines both a generalized phase space and an associated Fock-state lattice, and by exploiting how displacement operators act on the corresponding coherent states we obtain a natural realization of the walker dynamics. Within this framework we analyze DTQWs associated with several different algebras, leading to both one-dimensional and higher-dimensional walks. We show that the resulting dynamics typically exhibit ballistic spreading, but can also display richer behavior. In particular we identify a localization effect that appears when the displacement induces only small shifts of the walker between successive coin operations, and we present an example of an anomalous walk displaying super-ballistic spreading. Because the effective tunneling amplitudes on FSLs generally depend on occupation number, the lattices are typically not translationally invariant, and the resulting walks may be interpreted as occurring on effectively curved synthetic lattices.

The paper is organized as follows. In the next section we briefly review standard DTQWs, focusing on the one-dimensional case and summarizing several general properties. We also discuss implementations of DTQWs in phase space, which provide an important starting point for our construction. In Sec.~\ref{sec:FSLwalk} we introduce the general framework for DTQWs on Fock-state lattices. The formal construction is presented in Subsec.~\ref{ssub:gen_setting}, followed by a series of explicit examples in Subsec.~\ref{ssec:examples}. Finally, in Sec.~\ref{sec:con} we discuss possible experimental realizations and extensions of the scheme, and conclude.

\section{Background: Discrete-time quantum walks}

In order to develop walks on FSLs, we first review the
standard discrete-time quantum walk (DTQW) on a one-dimensional lattice~\cite{kempe2003quantum,venegas2012quantum}. This
serves to clarify the difficulty mentioned in the Introduction, namely that
implementing DTQWs generally requires engineered long-range Hamiltonians and
precise timing control.

\subsection{Regular DTQW on a one-dimensional lattice}
\label{ssec:1dDTQW}

We consider a single particle (the walker, $w$) moving on a one-dimensional
lattice. The state $|j\rangle$ denotes the walker localized at lattice site
$j\in\mathbb{Z}$. In addition, the walker possesses a two-dimensional internal
(coin) degree of freedom with basis states $|\!\!\uparrow\rangle$ and
$|\!\!\downarrow\rangle$. The Pauli matrices acting on the coin Hilbert space are
denoted by $\hat{\sigma}_\alpha$ ($\alpha=x,y,z$), e.g.,
$\hat{\sigma}_z|\!\!\uparrow\rangle=|\!\!\uparrow\rangle$ and
$\hat{\sigma}_x|\!\!\uparrow\rangle=|\!\!\downarrow\rangle$. The total Hilbert space is
\begin{equation}
    \mathcal{H}=\mathcal{H}_w\otimes\mathcal{H}_c.
\end{equation}

The discrete-time evolution consists of alternating applications of a coin
operator $\hat{U}_c$ (coin tossing) and a conditional shift operator $\hat{U}_w$. One step of
the walk is generated by
\begin{equation}
    \hat{W}=\hat{U}_w\hat{U}_c,
\end{equation}
so that after $m$ steps the state of the system is
\begin{equation}\label{sevolve}
    |\psi(m)\rangle=\hat{W}^m|\psi(0)\rangle.
\end{equation}
The dynamics is nontrivial because $\hat{U}_c$ and $\hat{U}_w$ do not commute.

For a two-sate coin we choose the coin operator to be the Hadamard transformation,
\begin{equation}\label{coinop}
    \hat{U}_c
    =\mathbb{I}_w\otimes\frac{1}{\sqrt{2}}
    \begin{pmatrix}
        1 & 1 \\
        1 & -1
    \end{pmatrix}.
\end{equation}
Since the Hadamard matrix satisfies $H^2=\mathbb{I}$ and has eigenvalues
$\pm1$, a convenient Hermitian generator is
\begin{equation}
    \hat{H}_c
    =\mathbb{I}_w\otimes\frac{\pi}{2}
    \left(
    \frac{1}{\sqrt{2}}
    \begin{pmatrix}
        1 & 1 \\
        1 & -1
    \end{pmatrix}
    \right),
\end{equation}
which yields $\hat{U}_c=e^{-i\hat{H}_c}$ up to a global phase.

The conditional shift operator entangles the walker and coin degrees of freedom
and is given by
\begin{equation}\label{wshift}
    \hat{U}_w
    =\sum_j\Big(
    |j+1\rangle\langle j|\otimes\up\langle\uparrow\!\!|
    +
    |j-1\rangle\langle j|\otimes\down\langle\downarrow\!\!|
    \Big).
\end{equation}
Thus, the walker moves one lattice site to the right (left) conditioned on the
coin being in state $|\!\!\uparrow\rangle$ ($|\!\!\downarrow\rangle$).

To derive the generator of the shift operator, we consider an infinite lattice
and exploit translational invariance. Introducing quasi-momentum states
\begin{equation}
    |k\rangle=\frac{1}{\sqrt{2\pi}}\sum_j e^{ikj}|j\rangle,
    \qquad k\in[-\pi,\pi],
\end{equation}
the shift operator becomes diagonal in momentum space,
\begin{equation}
    \tilde{U}_w(k)
    =
    e^{-ik}\up\langle\uparrow\!\!|
    +
    e^{ik}\down\langle\downarrow\!\!|
    =
    e^{-ik\hat{\sigma}_z}.
\end{equation}
The corresponding Hamiltonian can therefore be written as
\begin{equation}
\label{so}
    \hat{H}_w
    =
    \int_{-\pi}^{\pi}
    k\,|k\rangle\langle k|\,dk
    \otimes\hat{\sigma}_z.
\end{equation}
Transforming back to the site basis yields
\begin{equation}
\label{wham}
    \hat{H}_w
    =
    \sum_{j\neq l}
    \frac{(-1)^{j-l}}{i(j-l)}
    |j\rangle\langle l|
    \otimes\hat{\sigma}_z,
\end{equation}
where the sum is understood in the principal-value sense. Equation~(\ref{wham})
explicitly shows that the generator of the shift is long-ranged, with algebraic
decay $\sim1/|j-l|$.

Equation~(\ref{so}) has the form of a coin-conditioned translation: for the coin
state $\up$ the walker acquires a phase $e^{-ik}$, corresponding to
a unit translation in one direction, while for $\down$ the phase is
$e^{ik}$, corresponding to a translation in the opposite direction. This
observation will be important in the construction of shift operators on
FSLs in the following section.

The most prominent features of the one-dimensional DTQW
are~\cite{kempe2003quantum,venegas2012quantum}:
\begin{itemize}
    \item \textbf{Ballistic spreading.} The probability distribution $P_w(j,m)$
    spreads linearly in time, with width $\sigma_{\mathrm{qw}}\sim m$, in contrast
    to the diffusive scaling $\sigma_{\mathrm{cl}}\sim\sqrt{m}$ of a classical
    random walk.
    \item \textbf{Quantum interference.} The distribution is non-Gaussian and develops two counterpropagating peaks with interference fringes between them. 
    \item \textbf{Dependence on the initial coin state.} For a walker initially
    localized at the origin, symmetric spreading requires a complex superposition
    of coin states. For example, the state
    $(\up\rangle+i\down)/\sqrt{2}$ yields a symmetric
    distribution, whereas $(\up+\down)/\sqrt{2}$
    produces an asymmetric one.
    \item \textbf{Parity constraint.} After $m$ steps, the walker occupies only
    lattice sites $|j\rangle$ with $|j|\le m$ and parity matching that of $m$.
\end{itemize}

Finally, one may define a family of evolution operators parameterized by a
continuous variable $t$,
\begin{equation}
    \hat{W}(t)=e^{-i\hat{H}_w t}e^{-i\hat{H}_c t}.
\end{equation}
For $t\neq1$, the resulting probability distribution differs from that of the
standard DTQW. For $t<1$, the shift corresponds to a fractional lattice
translation in momentum space, and the coin rotation angle is reduced
accordingly. For even integers $t=2q$, the coin evolution satisfies
$e^{-i\hat{H}_c t}=(-1)^q\mathbb{I}$, so that the dynamics reduces to two
counterpropagating components up to a global phase. For $t<1$, the left/right shifted walker states have non-zero overlap. When this is the case, i.e. the walker states are non-orthogonal, and due to interferences the walker will in general not be symmetric given the initial coin state $(\up+i\down)/\sqrt{2}$ and walker state symmetric around $j=0$.


\subsection{DTQW in phase space}\label{ssec:phase_space_qw}

The long-range structure of the Hamiltonian in Eq.~(\ref{wham}) renders direct
implementations of DTQWs in discrete lattice models impractical. To circumvent this difficulty, quantum walks in phase space have been proposed and implemented~\cite{sanders2003quantum,xue2008quantum,xue2009quantum,schmitz2009quantum,zahringer2010realization,flurin2017observing}. The key observation is that coherent states of bosonic
modes are highly controllable in quantum optical platforms such as trapped ions and cavity QED systems. In particular, coherent states can be displaced in phase space while preserving their form.

We first consider a walk along a line in phase space~\cite{xue2009quantum,schmitz2009quantum,zahringer2010realization}, implemented through
successive applications of the displacement operator~\cite{mandel1995optical}
\begin{equation}\label{origdis}
    \hat{D}(\beta)=e^{\beta\hat{a}^\dagger-\beta^*\hat{a}},
\end{equation}
which acts on a Glauber coherent state $|\alpha\rangle$ as
\begin{equation}
    \hat{D}(\beta)|\alpha\rangle
    =
    e^{\frac{1}{2}(\beta\alpha^*-\beta^*\alpha)}
    |\alpha+\beta\rangle.
\end{equation}
Here $\hat{a}^\dagger$ and $\hat{a}$ denote bosonic creation and annihilation operators satisfying $[\hat{a},\hat{a}^\dagger]=1$.

Choosing $\alpha=0$ and real $\beta$, we define a coin-conditioned walker
unitary
\begin{equation}
    \hat{U}_w(\beta)
    =
    e^{(\beta\hat{a}^\dagger-\beta\hat{a})\hat{\sigma}_z}
    =
    \hat{D}(\beta)\otimes\up\langle\uparrow\!\!|
    +
    \hat{D}(-\beta)\otimes\down\langle\downarrow\!\!|.
\end{equation}
Starting from the initial state $|0\rangle\otimes\up$ and applying
a Hadamard coin followed by $\hat{U}_w(\beta)$, the state after one step is
\begin{equation}
    |\psi(1)\rangle
    =
    \frac{1}{2}
    \Big[
    (1+i)|+\beta\rangle\otimes\up
    +
    (1-i)|-\beta\rangle\otimes\down
    \Big],
\end{equation}
and after two steps one obtains
\begin{equation}
\begin{split}
|\psi(2)\rangle =
&\;\frac{1+i}{2\sqrt{2}}\,|+2\beta\rangle \otimes \up
- \frac{1-i}{2\sqrt{2}}\,|-2\beta\rangle \otimes \down \\
&+ \frac{1}{2}\,|0\rangle \otimes \big(\up + \down\big).
\end{split}
\end{equation}
The coefficients coincide with those of the standard DTQW on a
one-dimensional lattice. One may therefore identify the set of coherent states
$\{|l\beta\rangle\,:\,l\in\mathbb{Z}\}$ as effective lattice sites.

However, coherent states are not orthogonal. Their overlap is~\cite{mandel1995optical}
\begin{equation}\label{lover}
    \langle k\beta|l\beta\rangle
    =
    \exp\!\left[-\frac{\beta^2}{2}(k-l)^2\right],
    \qquad k,l\in\mathbb{Z},
\end{equation}
which is only exponentially suppressed. As a consequence, the resulting walker
probability distribution does not exactly reproduce that of the ideal DTQW.
Nevertheless, for sufficiently large $\beta$, the overlap becomes negligible
and the distributions approach one another.

An alternative to walking along the real axis of phase space is to perform the walk along a circle of radius $|\alpha_0|$~\cite{sanders2003quantum,xue2008quantum,flurin2017observing}. Let the initial walker
state be $|\beta_0\rangle$, with $\beta_0$ real. We define the walker unitary
\begin{equation}
\label{wuniphase}
    \hat{U}_w(\varphi)
    =
    e^{i\hat{n}\varphi\hat{\sigma}_z},
\end{equation}
where $\hat{n}=\hat{a}^\dagger\hat{a}$ is the boson number operator. Using
\begin{equation}
    e^{i\hat{n}\varphi}|\beta_0\rangle
    =
    |\beta_0 e^{i\varphi}\rangle,
\end{equation}
one obtains after two steps
\begin{equation}
\begin{split}
|\psi(2)\rangle =
&\;\frac{1+i}{2\sqrt{2}}\,
|\beta_0 e^{i2\varphi}\rangle \otimes \up
- \frac{1-i}{2\sqrt{2}}\,
|\beta_0 e^{-i2\varphi}\rangle \otimes \down \\
&+ \frac{1}{2}\,
|\beta_0\rangle \otimes \big(\up + \down\big).
\end{split}
\end{equation}
The overlap between two such phase-rotated coherent states is
\begin{equation}
\langle \beta_0 e^{ik\varphi} \,|\, \beta_0 e^{il\varphi} \rangle
=
\exp\!\left[
\!-|\beta_0|^2
\!+\! |\beta_0|^2 e^{i(l-k)\varphi}
\right]\!,
\quad k,l\in\mathbb{Z}.
\end{equation}
As in the linear case, orthogonality is recovered only in the limit of large
$|\beta_0|$ or sufficiently separated phase angles.

The unitary in Eq.~(\ref{wuniphase}) arises naturally in systems where a bosonic mode (such as a cavity field or a vibrational mode of a trapped ion)
interacts dispersively with a two-level system, leading to an effective
Hamiltonian proportional to $\hat{n}\hat{\sigma}_z$~\cite{larson2021jaynes}.


\section{Generalized discrete-time quantum walks: walks on Fock-state lattices}\label{sec:FSLwalk}

\subsection{General setting}\label{ssub:gen_setting}

Returning to DTQWs in phase space, either along a
line or on a circle, we note that the walker shift operator $\hat{U}_w$ is
expressed in terms of bosonic creation and annihilation operators
$\hat{a}^\dagger$ and $\hat{a}$. When working with bosonic operators, the most
natural basis is the Fock basis $\{|n\rangle\}$ with
$n\in\mathbb{Z}_{\geq0}$. For example, a coherent state may be expanded as
\cite{mandel1995optical}
\begin{equation}\label{gcoher}
    |\alpha\rangle
    =
    e^{-|\alpha|^2/2}
    \sum_{n=0}^{\infty}
    \frac{\alpha^n}{\sqrt{n!}}\,|n\rangle.
\end{equation}
The corresponding number distribution $P_\alpha(n)$ is Poissonian, with mean
$\langle\hat{n}\rangle=|\alpha|^2$ and variance
$\Delta n^2=|\alpha|^2$.

For a DTQW implemented along a line in phase space, the state after $m$ steps is entangled between coin and walker degrees of freedom, and in particular, the walker occupies coherent states $|l\alpha\rangle$ with $|l|\le m$. If $|\alpha|$ is sufficiently large, the overlap between neighboring coherent states is small, see~(\ref{lover}). 
The corresponding number distributions $P_{l\alpha}(n)$, for $k=l+1$, also have a small overlap, given by
\begin{equation}
    O_l
    =
    \sum_n
    \sqrt{P_{(l+1)\alpha}(n)\,P_{l\alpha}(n)}
    =
    e^{-|\alpha|^2/2},
\end{equation}
which is independent of $l$ and coincides with the coherent-state overlap.
Hence, the phase-space DTQW on a line may alternatively be interpreted as a walk
on a lattice whose sites correspond to Fock states $|n\rangle$. This structure
is referred to as a \emph{Fock-state lattice} (FSL).

Traditionally, a FSL is defined by expressing a Hamiltonian
$\hat{H}$ as a matrix in a number basis, which we generically refer to as a Fock basis. Examples include bosonic and fermionic Fock bases, lattice site states $|j\rangle$, or spin states $|S,m\rangle$ satisfying
$\hat{S}_z|S,m\rangle=m|S,m\rangle$ and $\hat{S}^2|S,m\rangle=S(S+1)|S,m\rangle$. In this representation, diagonal matrix elements $\hat{H}_{ii}$ correspond to onsite energies, while off-diagonal elements $\hat{H}_{ij}$ describe transitions between sites $i$ and $j$.

The site index $i$ of the FSL is general and may label sites in lattices of
arbitrary dimension. As a rule of thumb, the dimensionality of the FSL equals
the number of continuous degrees of freedom of the Hamiltonian minus the number
of continuous symmetries \cite{saugmann2023fock}, with notable
exceptions~\cite{ferraro2026algebraic}. Discrete degrees of freedom and symmetries play an
analogous role, leading to lattices with a finite number of sites along the
corresponding directions. The FSL construction is most useful for systems with
relatively few degrees of freedom, since for many-body systems the dimensionality
and geometry of the FSL quickly become prohibitively complex.

As an illustrative example, consider the driven harmonic oscillator
\begin{equation}
\label{dhoham}
    \hat{H}_{\mathrm{dho}}
    =
    \Delta\,\hat{a}^\dagger\hat{a}
    +
    \eta\left(\hat{a}^\dagger+\hat{a}\right).
\end{equation}
In the Fock basis, this Hamiltonian is tridiagonal, with tunneling amplitudes $J_n=\eta\sqrt{n}$. Apart from the lower boundary imposed by the vacuum $|0\rangle$, the $n$-dependence of $J_n$ causes the lattice to be non-translationally invariant. This feature will be relevant for DTQWs on FSLs.

A naive approach to defining a DTQW on a FSL would be to replace the real-space lattice states $|j\rangle$ in the shift unitary~(\ref{wshift}) with Fock states $|n\rangle$. However, this leads to a highly nonlocal walker Hamiltonian of the
form~(\ref{wham}). For example, in the bosonic case one finds
$\hat{H}_w=\hat{h}_w\otimes\hat{\sigma}_z$, with
\begin{equation}
\hat{h}_w
=
\sum_{n\neq m}
\frac{(-1)^{n-m}}{i(n-m)}
\frac{(\hat{a}^\dagger)^{n-m}}{\sqrt{n!\,m!}}
(\hat{n})_m\,\hat{a}^m,
\qquad n\ge m,
\end{equation}
where we introduced the falling factorial
$(\hat{n})_k=\hat{n}(\hat{n}-1)\cdots(\hat{n}-k+1)$ and
$(\hat{n})_0=\mathbb{I}$. Such Hamiltonians involve highly nonlinear operator products and are not readily implementable in experimental settings, motivating the search for alternative constructions.

To develop a more tractable approach, we adopt a Lie-algebraic perspective on FSLs, following Ref.~\cite{ferraro2026algebraic}. Consider a set of bosonic operators
\begin{equation}
    \{\hat{a},\hat{a}^\dagger,\hat{n},\mathbb{I}\}_{\mathfrak{hw}},
\end{equation}
which form the extended Heisenberg--Weyl algebra. The number operator $\hat{n}$
is a \emph{Cartan generator}, while $\hat{a}$ and $\hat{a}^\dagger$ are
\emph{root generators} appearing in Hermitian-conjugate pairs. The identity is
an \textit{ideal} commuting with all other generators~\cite{georgi2000lie}. In general,
the Cartan generators $\hat{C}_a$ are those that can be simultaneously
diagonalized; their eigenstates define the Fock basis. The number of Cartan
generators equals the rank $r$ of the algebra, and the corresponding eigenvalues
label the sites of the \emph{weight lattice}, whose dimension is $r$.

We define a (Lie) Fock-state lattice by identifying the sites of the weight
lattice with the corresponding Fock states. The connections between sites are
determined by the root generators $\hat{E}_{\pm\alpha}$. If there are $n$ root
pairs, then each site couples to $2n$ neighboring sites, with the root vectors
$\alpha$ determining the directions and distances of these couplings. For the
Heisenberg--Weyl algebra, there is a single Cartan generator $\hat{C}_1=\hat{n}$ and one
root pair $\hat{E}_{+1}=\hat{a}^\dagger$ and $\hat{E}_{-1}=\hat{a}$, resulting in a one-dimensional FSL
with tunneling amplitudes $J_n=\sqrt{n}$. This geometry coincides with that of
the Hamiltonian~(\ref{dhoham}), although the Lie-algebraic construction itself
is purely mathematical and does not include physical parameters such as
$\Delta$ or $\eta$.

An additional ingredient required for constructing DTQWs is the existence of a
reference state $|\psi_0\rangle$ that is annihilated by all negative root
generators,
\begin{equation}
    \hat{E}_{-\alpha}|\psi_0\rangle=0,
    \qquad \forall\,\alpha.
\end{equation}
One then defines displacement operators~\cite{georgi2000lie}
\begin{equation}
\label{gendis}
    \hat{D}_\alpha(\beta)
    =
    e^{\beta\hat{E}_{+\alpha}-\beta^*\hat{E}_{-\alpha}},\quad\beta\in\mathbb{C},
\end{equation}
and the associated generalized or Perelomov coherent states~\cite{perelomov1977generalized}
\begin{equation}
    |\beta\rangle
    =
    \hat{D}_\alpha(\beta)|\psi_0\rangle.
\end{equation}
For a single root pair $\hat{E}_{\pm\alpha}$ these cover all coherent states, while for algebras with multiple roots, the generator of a general coherent state takes the form
\begin{equation}\label{coherentgenerator}
    \hat{\Omega}(\beta)
    = e^{\sum_i\left(\beta_i \hat{E}_{+\alpha_i} -  \mathrm{H.c.}\right)}, \qquad \beta_i \in \mathbb{C},
\end{equation}
since the root generators do not mutually commute. We may mention that not every algebra supports a reference state, especially if the FSL is unbounded. 

The set of Perelomov coherent states forms a manifold, which may be interpreted
as a Lie-algebraic phase space (LPS). On this manifold one can define generalized
Husimi functions for a state $\hat{\rho}$,
\begin{equation}
    Q(\beta)
    =
    c(\beta)\langle\beta|\hat{\rho}|\beta\rangle,
\end{equation}
where $c(\beta)$ is a normalization factor. If $\hat{\rho}$ itself is a
Perelomov coherent state, then $Q(\beta)$ is localized and (approximately)
Gaussian on the LPS and represents a minimum-uncertainty state with respect to the algebra generators. Acting repeatedly with $\hat{D}_\alpha(\beta)$ displaces the state along a geodesic of the LPS. In general, the distance along this geodesic is determined by the invariant metric of the manifold and need not equal $|\beta|$. For several commonly encountered algebras, the displacement distance is given by
\begin{equation}
\label{psdist}
\begin{aligned}
    d_{\mathfrak{e}(2)} &= |\beta|, \\
    d_{\mathfrak{hw}} &= |\beta|, \\
    d_{\mathfrak{su}(2)} &= \arccos\!\left[\cos^{2S}(\beta)\right], \\
    d_{\mathfrak{su}(1,1)} &= \arccos\!\left[(1-|\beta|^2)^k\right],
\end{aligned}
\end{equation}
where $\mathfrak{e}(2)$ denotes the Euclidean algebra and $\mathfrak{hw}$ the
Heisenberg--Weyl algebra.

As an illustrative example, consider $\mathfrak{su}(2)$, the algebra of angular
momentum, generated by
\begin{equation}\label{su2elements}
    \{\hat{S}_x,\hat{S}_y,\hat{S}_z\}_{\mathfrak{su}(2)}.
\end{equation}
The Cartan generator is $\hat{S}_z$, with eigenvalues $-S\le l\le S$ and
corresponding Fock states $|S,l\rangle$. The resulting FSL is therefore a finite
one-dimensional chain. The \textit{Casimir operator},
\begin{equation}\label{su2casimir}
    \hat{S}^2
    =
    \hat{S}_x^2+\hat{S}_y^2+\hat{S}_z^2
    =
    S(S+1),
\end{equation}
defines the geometry of the LPS, which is a sphere. While one might expect the
radius to be $\sqrt{S(S+1)}$, the correct radius is $S$, reflecting the fact that
for spin coherent states $\langle\hat{\mathbf{S}}\rangle^2=S^2$.

Having introduced the necessary Lie-algebraic framework, we now outline how to
construct DTQWs on FSLs. The key idea is that, in many cases, a state localized
on the LPS also corresponds to a localized state on the FSL. One may therefore
define a quantum walk on the LPS using the displacement operators
(\ref{gendis}), and subsequently project the resulting dynamics onto the FSL.
The walker shift operator is taken as
\begin{equation}\label{walkerun}
    \hat{U}_w(\beta)
    =
    e^{(\beta\hat{E}_{+\alpha}-\beta^*\hat{E}_{-\alpha})\otimes\hat{\sigma}_z},
\end{equation}
with the displacement conditioned on the coin state. As before, the coin operator is chosen to be the Hadamard. The parameter $\beta$ plays the role of a step size, and because Perelomov coherent states are nonorthogonal, it controls the distinguishability of different walker states.

The full state then evolves according to Eq.~(\ref{sevolve}), with
\begin{equation}\label{fullun}
    \hat{W}(\beta) = \hat{U}_w(\beta)\hat{U}_c.
\end{equation}
After $m$ steps, the effective `time' is $T = m\beta$, and the full evolution operator is
\begin{equation}
    \hat{U}(T) = \hat{W}^m(\beta).
\end{equation}
One can explore the Trotterized continuum limit $\beta \to 0$ while keeping $T$ fixed, which corresponds to continuous-time evolution. Expanding $\hat{U}_w(\beta)$ to first order in $\beta$ yields
\begin{equation}
\label{trotter}
    \hat{U}(T) \approx \hat{U}_c^m\, e^{-i \hat{H}_\mathrm{eff} T},
\end{equation}
with the effective Hamiltonian
\begin{equation}\label{contham}
    \hat{H}_\mathrm{eff} = i\left(\hat{E}_{-\alpha}-\hat{E}_{+\alpha}\right)\otimes \hat{\sigma}_x.
\end{equation}
Here, we may interpret
\begin{equation}
    \hat{X}_\alpha = i\left(\hat{E}_{-\alpha}-\hat{E}_{+\alpha}\right)
\end{equation}
as a momentum operator. The fast-oscillating unitary
$\hat{U}_c^m$ does not have a well-defined limit as $m\to\infty$. Nevertheless, Eq.~(\ref{trotter}) shows that for small $\beta$ the evolution produces two phase-space distributions moving in opposite directions, one associated with spin-up and one with spin-down along the $x$ axis. The rapid coin oscillations effectively average the populations, ensuring that the two wave packets carry equal weight. However, for small but non-zero $\beta$, the shifted walker states will greatly overlap causing not only destructive interference, but also an asymmetric walk even for the initial coin state $(\up+i\down)/\sqrt{2}$ and initial symmetric walker state. This is of importance for us since there will always be a slight overlap between the Perelomov coherent states, see for example~(\ref{lover}), that will result in small asymmetries in the walker distribution.

Finally we note that a crucial aspect of this construction is the projection of the phase-space
distribution onto the FSL. While this becomes difficult to visualize in higher dimensions, it is transparent in the $\mathfrak{su}(2)$ case. There, the LPS is the Bloch sphere and the Perelomov coherent states are the familiar spin coherent states~\cite{mandel1995optical}. The mapping to the FSL distribution is obtained by projecting the Husimi distribution onto the $z$ axis, corresponding to the FSL generated by the Cartan operator $\hat{S}_z$. We demonstrate this mapping explicitly in the following section.


\subsection{Examples}\label{ssec:examples}
In this section we consider some of the most familiar Lie algebras and apply the idea of a DTQW on their FSLs. Thus, given an algebra we identify its LPS and FSL, as well as the reference state and displacement operators. After that we explore the properties of the corresponding DTQW.

\begin{figure}[!ht]
    \centering
    \includegraphics[width=0.7\linewidth]{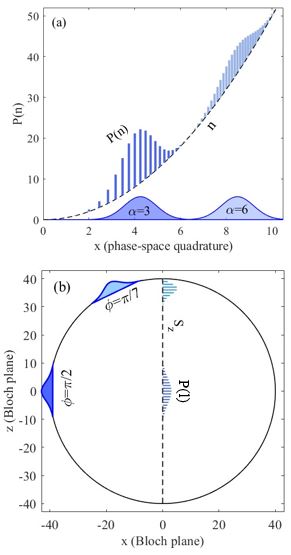}
    \caption{Schematic illustration of the LPS--FSL correspondence for the Heisenberg-Weyl (a) and $\mathfrak{su}(2)$ (b) models. For the $\mathfrak{hw}$ model, the LPS is the $x$-$p$ plane. A coherent state with real amplitude $\alpha$ has a Gaussian phase-space distribution with mean momentum quadrature $p = 0$ and $x = \sqrt{2}\,\alpha$; the marginal $|\psi(x)|^2$ is shown for $\alpha = 3$ and $6$. The diagonal Cartan generator is $\hat{n}$, so that the FSL corresponds to the $y$ axis. To connect the LPS to the FSL, we consider the parabolic dashed curve/surface $n = (x^2+p^2)/2$; the boson probability distributions $P(n)$ are obtained by first projecting the Gaussian onto this curve and then onto the $n$ axis perpendicular to the LPS. For the $\mathfrak{su}(2)$ model of a spin-$S$ system, the phase space is the Bloch sphere of radius $S$ (here $S = 40$), and the Cartan operator $\hat{S}_z$ is represented by the $z$ axis. Spin-coherent states move along the great circle with polar angle $\theta = \pi/2$ (the $x$-$z$ plane), and projecting onto this axis gives the FSL probability distributions $P(l)$, shown for $\phi = \pi/2$ (equator) and $\phi = \pi/7$. As the phase-space distribution approaches the poles, the variance of the corresponding $P(l)$ distributions narrow, vanishing exactly at the poles. In both plots, and all other 1D FSL examples in this paper, the initial coin state is $(\up+i\down)/\sqrt{2}$.}
    \label{fig:ps_fsl}
\end{figure}


\subsubsection{Heisenberg-Weyl algebra}
The $\mathfrak{hw}$ algebra and its properties were introduced above: the LPS is the regular $x$-$p$ phase space, the coherent states are the standard Glauber coherent states~(\ref{gcoher}), the displacement operator was given in Eq.~(\ref{origdis}), and the FSL is a 1D chain with sites labeled by $n\in\mathbb{Z}_{\geq 0}$ and tunneling amplitudes $J_n=\sqrt{n}$. Since the lattice sites are labeled by $n$, which is related to the squared amplitude of the coherent state, a fixed $n$ in phase space corresponds to a circle of radius $n = (x^2+p^2)/2$. 

We may visualize this by introducing a third axis, $n$, perpendicular to the $x$-$p$ plane, so that each axis represents one generator of the algebra. To connect the FSL (along the $n$ axis) to the LPS, we consider the parabolic surface $(x^2+p^2)/2$. A Husimi function $Q(\alpha)$ on the LPS can then be projected onto this surface, followed by a second projection down to the $n$ axis to obtain the boson distribution $P(n)$, as illustrated in Fig.~\ref{fig:ps_fsl}(a). In the figure we show a slice of phase space with two coherent states, $\alpha = 3$ and $6$, along with the corresponding $P(n)$ distributions projected onto the parabolic surface. The widths of the Husimi functions are independent of $\alpha$, whereas the widths of $P(n)$ are not. Moreover, the distance between consecutive distributions (associated with displaced coherent states) varies along the FSL depending on the coherent state amplitude.

Turning to the DTQW, consider displacements along the $x$ axis of phase space, so that the walker shift operator can be written as
\begin{equation}
    \hat{U}_w(\beta) = e^{i\sqrt{2}\,\beta \,\hat{p}\otimes \hat{\sigma}_z},
\end{equation}
with $\beta$ real and $\hat{p} = i(\hat{a}^\dagger - \hat{a})/\sqrt{2}$ the
$p$-quadrature of the boson mode. For the phase-space DTQW along a line (subsec.~\ref{ssec:phase_space_qw}), it was natural to initialize the walker in the vacuum state. For a walk on the FSL, however, this scheme does not produce the desired coin-conditioned displacement. Instead, one should initialize the walker in a coherent state $|\alpha\rangle$ with $\alpha$ real and sufficiently large so that displaced states remain on the positive $x$ axis, avoiding the $n=0$ boundary. 

\begin{figure*}[!ht] 
    \centering
    \includegraphics[width=0.78\textwidth]{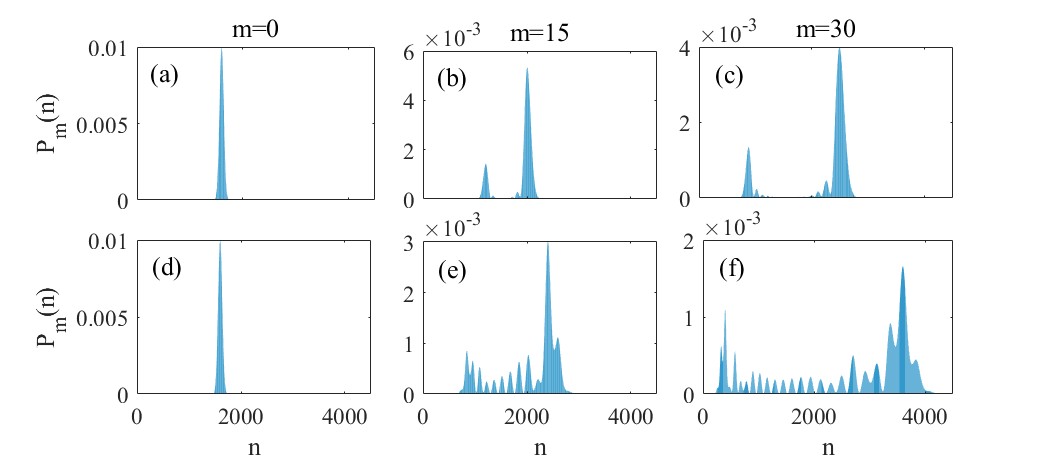} 
    \caption{Two examples of the boson number distribution $P_m(n)$ on the Fock-state lattice of the $\mathfrak{hw}$ algebra, after $m$ time steps. The initial state is a coherent state with $\alpha = 40$, corresponding to $\langle \hat{n} \rangle = 1600$. In the upper panel the displacement size is $\beta = 1/2$, while in the lower panel $\beta = 1$. For the smaller $\beta$, the displaced coherent states overlap, producing destructive interference near the center of the distribution (upper panel). For the larger $\beta$, the coherent states are approximately well separated, resulting in the characteristic interference pattern of a discrete-time quantum walk (lower panel). }
    \label{fig:hw}
\end{figure*}

Numerical results for the distributions $P_m(n)$ for two such simulations are shown in Fig.~\ref{fig:hw}, after three different number of steps and for two different displacements $\beta$. For the smaller $\beta$, consecutive coherent states overlap, producing destructive interference in the center of the walker distribution, in agreement with the analytical continuum-limit result~(\ref{trotter}).  Increasing $\beta$ reduces this overlap and restores the characteristic DTQW interference pattern.

Finally, we verified that the width of the distribution scales ballistically, $\sigma_\mathfrak{hw} = \Delta n \sim m \beta$, as expected. Another hallmark of DTQWs is the rapid buildup of coin-walker entanglement~\cite{abal2006quantum,venegas2012quantum}.  Although not shown here, we observe a rapid increase in entanglement that then remains high. In the continuous-time limit~(\ref{trotter}), this behavior can be understood in terms of the two separating wave packets: once they cease to overlap, the coin becomes maximally entangled with the walker. This phenomenon is reminiscent of the preparation of Schr\"odinger cat states~\cite{brune1992manipulation,brune1996observing}, 
though the preparation method here is different.
   

\subsubsection{$\mathfrak{su}(2)$ algebra}
Some properties of the $\mathfrak{su}(2)$ algebra were already mentioned in subsec.~\ref{ssub:gen_setting}, in particular how the Casimir operator ~(\ref{su2casimir}) determines the LPS. Like for the $\mathfrak{hw}$ algebra, the spin-coherent states, the spin phase space, and displacements on this space are standard material covered in textbooks~\cite{klauder1985coherent,
gazeau2009coherent,klimov2009group}. 

The elements of the algebra, given in~(\ref{su2elements}), obey the standard angular momentum commutation relations,
\begin{equation}
    [\hat{S}_i,\hat{S}_j] = i \varepsilon_{ijk} \hat{S}_k,
\end{equation}
where $\varepsilon_{ijk}$ is the Levi-Civita tensor. Equivalently, one can work with the ladder operators,
\begin{equation}
    [\hat{S}^+,\hat{S}^-] = 2 \hat{S}_z, \quad [\hat{S}_z,\hat{S}^\pm] = \pm \hat{S}^\pm,
\end{equation}
which act on the Fock states $|S,l\rangle$ as
\begin{equation}\label{spinfac}
    \hat{S}^\pm |S,l\rangle = \sqrt{S(S+1)-l(l\pm 1)}\,|S,l\pm1\rangle.
\end{equation}

\begin{figure}[!ht] 
    \centering
    \includegraphics[width=0.95\linewidth]{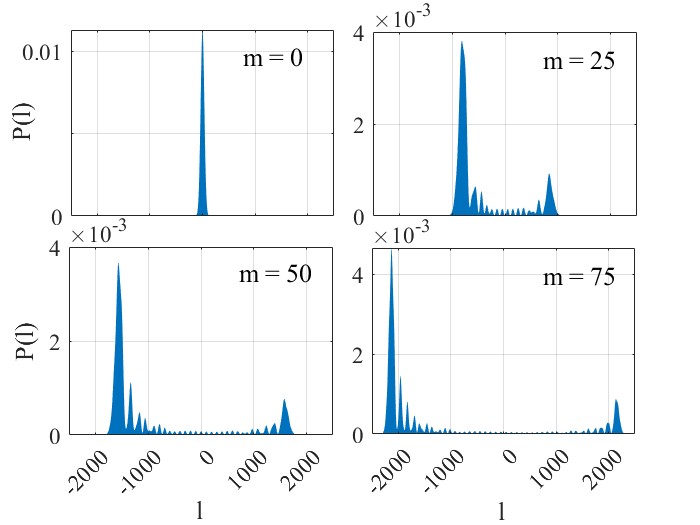} 
    \caption{Snapshots of the distribution $P(l)$ for the DTQW on the $\mathfrak{su}(2)$ Fock-state lattice. The initial walker state is the spin-coherent state $|\theta = \pi/2, \phi = \pi/2\rangle$, so that the Husimi function is localized on the equator pointing along the positive $y$ direction, producing a distribution centered in the FSL (upper-left plot, $m = 0$). As the walk evolves, the shift occurs along the $\theta = \pi/2$ meridian, broadening $P(l)$; after $m = 75$ steps (lower-right plot), the walker nearly reaches the boundaries of the FSL ($S = 2500$). When the FSL boundaries are approached, the Husimi function extends to the north and south poles. The asymmetry of the distribution arises because spin-coherent states are not perfectly orthogonal. Here, the step size is $\zeta = 0.01i$, giving a clear non-vanishing overlap between the coherent states and hence an asymmetric walker distribution.}
    \label{fig:su2}
\end{figure}

The Bloch-sphere LPS can be parametrized by polar and azimuthal angles $\theta\in[0,\pi]$ and $\phi\in[0,2\pi)$. Defining the complex displacement parameter
\begin{equation}
    \zeta = -\frac{\theta}{2} e^{i\phi},
\end{equation}
the spin displacement operator is~\cite{zhang1990coherent}
\begin{equation}\label{spindisplace}
    \hat{D}(\zeta) = e^{\zeta \hat{S}^+ - \zeta^* \hat{S}^-}.
\end{equation}
With reference state $|S,S\rangle$, the spin-coherent states are
\begin{equation}
    |\theta,\phi\rangle = \hat{D}(\zeta) |S,S\rangle.
\end{equation}
Rotations $R_i(\varphi) = e^{i\varphi \hat{S}_i}$ ($i=x,y$) are special realizations of displacements along the meridians, while $R_z(\varphi) = e^{i\varphi \hat{S}_z}$ shifts the azimuthal angle along parallels of the Bloch sphere. The Husimi function of a spin-coherent state $|\theta_0,\phi_0\rangle$ is
\begin{equation}\label{su2husimi}
    Q(\theta,\phi) = |\langle \theta,\phi|\theta_0,\phi_0\rangle|^2
    = \left[\cos\frac{\gamma}{2}\right]^{4S},
\end{equation}
where $\gamma$ is the angular distance between $(\theta,\phi)$ and $(\theta_0,\phi_0)$, related to the geodesic distance $d$ via $\gamma = d/S$:
\begin{equation}
    \cos \gamma = \cos\theta \cos\theta_0 + \sin\theta \sin\theta_0 \cos(\phi-\phi_0).
\end{equation}
As expected for a Perelomov coherent state, $Q(\theta,\phi)$ is localized around $(\theta_0,\phi_0)$. Using the composition formula $\hat{D}(\xi)\hat{D}(\zeta) = e^{i\Phi(\xi,\zeta)}\hat{D}(\varsigma)$ for some phase $\Phi(\xi,\zeta)$, it follows that coherent states remain coherent under successive displacements.

The relation between LPS and FSL distributions is
\begin{equation}
\begin{aligned}
    |\theta,\phi\rangle &= \sum_{l=-S}^{S} \binom{2S}{S+l}^{1/2}
    \left(\cos\frac{\theta}{2}\right)^{S+l} \left(\sin\frac{\theta}{2}\right)^{S-l} \\
    &\quad \times e^{-i(S-l)\phi}\, |S,l\rangle.
\end{aligned}
\end{equation}
The Cartan generator $\hat{S}_z$ defines the FSL $z$-axis, and the Husimi distributions are projected onto this axis to obtain $P(l)$, as illustrated in fig.~\ref{fig:ps_fsl}(b). Coherent states on the equator ($\phi = \pi/2$) produce wide $P(l)$ distributions at the center of the FSL, while states near the poles generate narrower $P(l)$ distributions close to the FSL boundaries.

For the DTQW on the $\mathfrak{su}(2)$ FSL, we consider rotations around the $x$ axis using a purely imaginary $\zeta$. The initial walker is at the equator, so it starts at the center of the FSL. Snapshots of $P(l)$ are shown in
fig.~\ref{fig:su2} for $m = 0, 25, 50, 75$. As expected for a quantum walk, the spreading $\Delta l$ is ballistic, and the walker-coin entanglement grows rapidly. The FSL distributions are asymmetric due to non-orthogonality of spin-coherent
states. For small $|\zeta|$, the overlap between neighboring states is
\begin{equation}
    |\langle \theta + |\zeta|, \pi/2 | \theta, \pi/2 \rangle|^2
    \approx e^{-S|\zeta|^2/2},
\end{equation}
which decreases with increasing spin $S$. For the parameters in
fig.~\ref{fig:su2} ($S=2500$, $|\zeta|=0.01$), the overlap is $\approx 0.88$, producing a strongly asymmetric walker distribution; increasing, for example, $|\zeta| = 0.05$ the overlap is down to $0.04$, considerably reducing asymmetry.  


\subsubsection{Two-dimensional DTQW: $\mathfrak{su}(3)$ and $\mathfrak{so}(5)$}
For the previous examples we considered algebras supporting one-dimensional FSLs, in which case the coin has two internal states. In higher-dimensional FSLs there are additional tunneling terms, each associated with a distinct coin state~\cite{mackay2002quantum,grimmett2004weak,inui2004localization}. Here we present two examples of DTQWs on two dimensional FSLs based on the
$\mathfrak{su}(3)$ and $\mathfrak{so}(5)$ algebras.

The $\mathfrak{su}(3)$ Lie algebra contains eight generators: two Cartan generators and six root generators. Consequently, the FSL is two-dimensional, with three roots $\alpha_1$, $\alpha_2$, and $\alpha_3=\alpha_1+\alpha_2$. The six root directions generate six neighbors for each site (apart from edge and corner sites), giving the FSL a triangular geometry~\cite{ferraro2026algebraic}. The minimal representation of the generators is provided by the $3\times3$ Gell-Mann matrices. For general representations it is convenient to use a bosonic construction applicable to any $\mathfrak{su}(n)$ algebra~\cite{anishetty2009irreducible}. For $\mathfrak{su}(3)$ we write
\begin{equation}
\begin{aligned}
    \hat{C}_1 &= \hat{n}_1-\hat{n}_2, \\
    \hat{C}_2 &= \frac{1}{\sqrt{3}}\left(\hat{n}_1+\hat{n}_2-2\hat{n}_3\right), \\
    \hat{E}_{+\alpha_1} &= \hat{a}_1^\dagger\hat{a}_2,\quad
    \hat{E}_{+\alpha_2} = \hat{a}_2^\dagger\hat{a}_3,\quad
    \hat{E}_{+\alpha_3} = \hat{a}_1^\dagger\hat{a}_3,
\end{aligned}
\end{equation}
with $\hat{E}_{-\alpha_i}=(\hat{E}_{+\alpha_i})^\dagger$. The total boson number $N=\hat{n}_1+\hat{n}_2+\hat{n}_3$ is a Casimir operator; fixing $N$ therefore determines the size of the FSL, which contains $(N+1)(N+2)/2$ sites. Along each principal axis of the triangular FSL one boson number is fixed, so the corner sites correspond to the Fock states $|N,0,0\rangle$, $|0,N,0\rangle$, and $|0,0,N\rangle$~\cite{saugmann2023fock}. The FSL distribution may be expressed either in terms of Cartan indices $(m_1,m_2)$ or the boson numbers $P(n_1,n_2,n_3)$, with the latter being more natural for numerical analysis. Although the LPS of $\mathfrak{su}(3)$ is four-dimensional and compact, its high dimensionality makes visualization of the LPS--FSL mapping difficult; we therefore focus on the FSL dynamics.

A reference state may be chosen as any corner state, from which generalized
Perelomov (spin) coherent states are obtained~\cite{perelomov1977generalized,
Perelomov1986,mathur2001coherent}:
\begin{equation}\label{su3cohe}
    |\beta\rangle \propto
    \left(\beta_1\hat{a}_1^\dagger+\beta_2\hat{a}_2^\dagger+\beta_3\hat{a}_3^\dagger
    \right)^N |0,0,0\rangle,
\end{equation}
with $\beta_i\in\mathbb{C}$ and $\sum_i|\beta_i|^2=1$. If $|\beta_i|=1$ the state coincides with a corner of the FSL, whereas $|\beta_i|=1/\sqrt{3}$ corresponds to the FSL center. The walker unitary is
\begin{equation}
    \hat{U}_w(\beta)
    =
    \sum_{i=1}^3\Big[
    \hat{D}_{\alpha_i}(\beta)|i\rangle\langle i|
    +
    \hat{D}_{\alpha_i}(-\beta)|i+3\rangle\langle i+3|
    \Big],
\end{equation}
where $|i\rangle$ ($i=1,\ldots,6$) denote the six coin states. Because the
walker employs a multi-state coin, we replace the two-state Hadamard operator with \textit{Grover’s diffusion operator}~\cite{shenvi2003quantum,
carneiro2005entanglement},
\begin{equation}
    \hat{U}_c = 2|s\rangle\langle s|-\mathbb{I},
\end{equation}
where $|s\rangle=(1,1,1,1,1,1)^T/\sqrt{6}$. This operator is also suitable for Grover’s search algorithm, and is only one of many possible ones. We initialize the coin in state $|s\rangle$ and the walker in the symmetric coherent state~(\ref{su3cohe}) with
$\beta_i=1/\sqrt{3}$, which ensures a symmetric walk.

\begin{figure}[!ht] 
    \centering
    \includegraphics[width=0.75\linewidth]{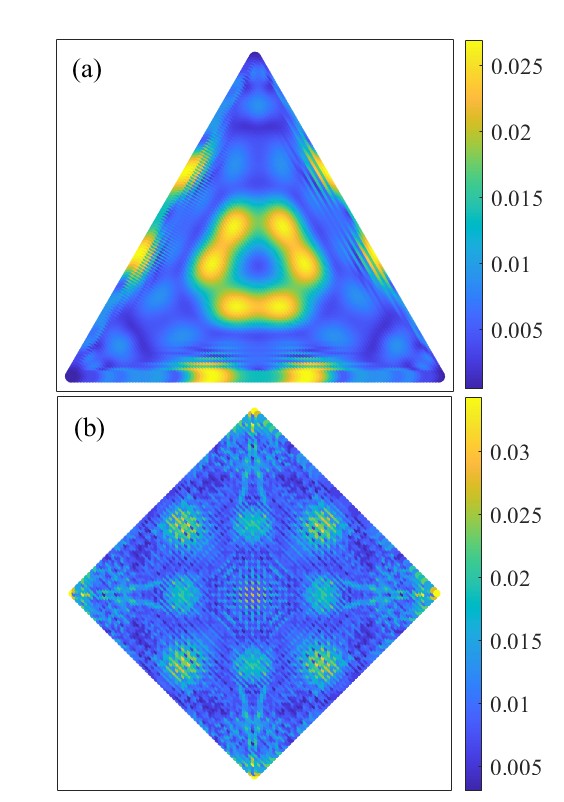} 
    \caption{Square root of the FSL distributions after five ($m = 5$) time steps for the $\mathfrak{su}(3)$ (a) and $\mathfrak{so}(5)$ (b) models. The total Hilbert-space dimensions are $3(N+1)(N+2) \sim 3N^2$ and $4(N^2+2N+2) \sim 4N^2$, respectively. Consequently, the $N^2$ scaling makes numerical simulations limited to moderate system sizes, and for the chosen step size $\beta$ the walker reaches the FSL boundaries after only a few steps. In both cases, the underlying lattice symmetry is clearly reflected in the interference structure of the distributions. As an initial walker state we consider the Perelomov coherent state in (a), and a Gaussian state in (b) (which approximates the Perelomov coherent state for a width $\sigma=\sqrt{N}/2$), and for both they are centered in the middle of the FSL. The initial coin state is the symmetric Grover state. The parameters are $N = 120$ and $\beta = 0.2$. }
    \label{fig:su3so5}
\end{figure}

Figure~\ref{fig:su3so5}(a) shows the distribution $P(n_1,n_2,n_3)$ after five steps. For the chosen parameters ($N=120$, $\beta=0.2$) the walker reaches the FSL boundaries after only a few steps, but the interference pattern and the sixfold rotational symmetry $C_6$ of the triangular lattice remain clearly visible.

Grover DTQWs on higher-dimensional lattices are known to exhibit both
ballistic and localized components; for example, if the walker starts at the
center site, that site retains non-negligible population at all times due to
dispersionless (flat) bands in the spectrum~\cite{mackay2002quantum,
inui2004localization}. Thus, such localization arises because the unitary evolution supports eigenvalues independent of momentum. In the FSL DTQW this effect is absent: the center of the lattice becomes almost empty after five steps. This difference is expected, since the FSL lacks exact translational symmetry and therefore does not support perfect flat bands, allowing dispersion of the walker.

\begin{figure*}[!ht] 
    \centering
    \includegraphics[width=0.6\textwidth]{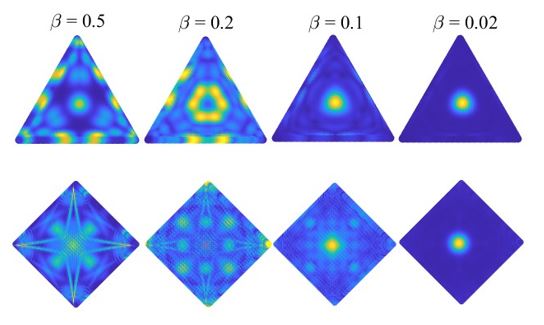} 
    \caption{Square roots of the walker distributions after time $T = m\beta = 1$ for the $\mathfrak{su}(3)$ model (upper panel) and the $\mathfrak{so}(5)$ model (lower panel), shown for different shift amplitudes $\beta = 0.5,\,0.2,\,0.1,\,0.02$ (from left to right). Importantly, in all cases the final evolution time of the DTQW is the same. As $\beta$ is reduced, and the kicking frequency of the coin is correspondingly increased, the walker distribution becomes increasingly localized, demonstrating dynamical localization. The system size is identical to that of Fig.~\ref{fig:su3so5}, namely $N = 120$.}
    \label{fig:localization}
\end{figure*}

Turning to the $\mathfrak{so}(5)$ algebra and its bosonic representation, we introduce two Cartan generators
\begin{equation}
\begin{aligned}
    \hat{H}_1 &= \frac{1}{2}\left(\hat{a}_\uparrow^\dagger\hat{a}_\uparrow
    - \hat{a}_\downarrow^\dagger\hat{a}_\downarrow\right), \\
    \hat{H}_2 &= \frac{1}{2}\left(\hat{b}_\uparrow^\dagger\hat{b}_\uparrow
    - \hat{b}_\downarrow^\dagger\hat{b}_\downarrow\right),
\end{aligned}
\end{equation}
and three pairs of root generators with positive roots
\begin{equation}
\begin{aligned}
    \hat{\Sigma}_{\alpha_1} &= \hat{a}_\uparrow^\dagger\hat{a}_\downarrow, \qquad
    \hat{\Sigma}_{\alpha_2} = \hat{b}_\uparrow^\dagger\hat{b}_\downarrow, \\
    \hat{\Sigma}_{\alpha_1+\alpha_2} &= \hat{a}_\uparrow^\dagger\hat{b}_\downarrow, \qquad
    \hat{\Sigma}_{\alpha_1+2\alpha_2} = \hat{a}_\downarrow^\dagger\hat{b}_\uparrow.
\end{aligned}
\end{equation}
Four bosonic modes are therefore required; the labels $a$ and $b$ with subscripts are chosen for notational convenience. The first two root generators describe nearest-neighbor tunneling, whereas the latter two generate next-nearest-neighbor tunneling. Consequently, the FSL is a square lattice with both axial and diagonal bonds. As in previous examples, the conservation of total boson number
\begin{equation}
    N = \hat{n}_{a_\uparrow}+\hat{n}_{a_\downarrow}
    +\hat{n}_{b_\uparrow}+\hat{n}_{b_\downarrow}
\end{equation}
is a Casimir operator and limits the size of the FSL. Using Cartan labels $(m_1,m_2)$, the FSL assumes the shape of a symmetric diamond; corner sites correspond to Fock states with only one mode populated. The LPS of $\mathfrak{so}(5)$ is compact and six-dimensional. Unlike the $\mathfrak{su}(3)$ case, Perelomov coherent states for this algebra are relatively involved, so we instead initialize the walker with a Gaussian distribution
\begin{equation}
    P(m_1,m_2) \propto
    \exp\!\left[-\frac{m_1^2+m_2^2}{\sigma^2}\right],
\end{equation}
with width $\sigma=\sqrt{N}/2$, which approximates the spread of a Perelomov coherent state.

The resulting numerical distribution is shown in fig.~\ref{fig:su3so5}(b). The Grover walk with the chosen coin state preserves the symmetry of the lattice, and (although not displayed) both the $\mathfrak{su}(3)$ and
$\mathfrak{so}(5)$ walks exhibit ballistic spreading.

Earlier we derived the continuum limit of the evolution operator for $\beta\rightarrow0$ at fixed $T=m\beta$, obtaining Eq.~(\ref{trotter}), in which the walker splits into left- and right-propagating components analogous to a
cat state. For the present two-dimensional multi-state walk, the same continuum procedure leads to a different result: the walker localizes rather than splitting. This is demonstrated in fig.~\ref{fig:localization} for both models. Keeping $T$ fixed while decreasing $\beta$, we observe that the full
walker distribution freezes as $\beta\rightarrow0$. This differs from Grover-type localization in higher dimensions, where a portion of the wave function populates dispersionless bands and only part of the state localizes; here the entire state becomes localized.

The coin unitary may be interpreted as periodic kicks of the coin subsystem, and such driving is known to produce dynamical localization in certain settings~\cite{fishman1982chaos,haake1991quantum}. In the kicked-rotor problem, the classical counterpart exhibits chaotic diffusion, whereas the quantum model localizes above a threshold kick strength. The phenomenon is understood by mapping the problem in Floquet (momentum) space to a one-dimensional Anderson model, which supports spatial localization. Although our system involves periodic kicks, the observed localization occurs in the limit $\beta\rightarrow0$, independent of the kick strength, so the analogy to the standard kicked-rotor mechanism is limited.

Instead, note that keeping $T=m\beta$ fixed while reducing $\beta$ requires more frequent applications of the coin unitary $\hat{U}_c$. Thus the coin evolves on a faster time scale than the walker, motivating a high-frequency Magnus expansion. Averaging over the rapidly oscillating coin degrees of freedom replaces the projectors $\hat{P}_i=|i\rangle\langle i|$ by $\mathbb{I}/d$, where $d$ is the coin dimension. Expanding the walker unitary $\hat{U}_w(\beta)$ for small $\beta$, the linear terms in the root generators
$\hat{E}_{\pm\alpha_i}$ cancel in this continuum limit for the Grover walk. This contrasts with the one-dimensional Hadamard walk, where the effective continuum Hamiltonian is linear in the momentum operator and generates ballistic counterpropagating wave packets. In two dimensions, second-order terms proportional to
$\frac{\beta^2}{2}[\hat{E}_{\pm\alpha_i},\hat{E}_{\pm\alpha_j}]$ survive because the Lie-algebra commutators are nonzero, but their contribution is suppressed by the $\beta^2$ prefactor. Consequently, spreading persists but with an effective velocity scaling as $v\sim\beta$ (rather than $\mathcal{O}(1)$ as in the one-dimensional case). The propagation is therefore ballistic but considerably slower, consistent with the observed localization in the
$\beta\rightarrow0$ limit.


\subsubsection{Noncompact Lie algebras and unconventional quantum walks on FSLs}

Following the general scheme developed above, we now present examples in which the resulting walks behave conceptually differently from standard DTQWs. The origin of this difference is not the same in the two models considered here.

We begin with the Euclidean Lie algebra $\mathfrak{e}(2)$~\cite{klimov2009group}, which at first sight resembles the Heisenberg--Weyl algebra. However, the two are fundamentally different. The generators satisfy
\begin{equation}
    [\hat{E}_0,\hat{E}^\pm] = \pm \hat{E}^\pm, \qquad
    [\hat{E}^-,\hat{E}^+] = 0,
\end{equation}
where $\hat{E}_0$ is the Cartan generator and $\hat{E}^\pm$ are root generators. While the first commutator mirrors the $\mathfrak{hw}$ case, the second differs crucially. In particular, there is no lowest (vacuum) state; the Fock states $|j\rangle$ span all integers $j\in\mathbb{Z}$. The algebra is therefore noncompact and admits no natural reference state from which to construct an LPS in the usual way. Nevertheless, since the root generators commute, the model possesses translational symmetry. Therefore, one may view the natural LPS as a cylinder, whose central axis represents the FSL~\cite{ferraro2026algebraic}.

In the Fock basis $|j\rangle$, the generators can be written as
\begin{equation}\label{e2gen}
    \hat{E}_0 = \sum_j j\,|j\rangle\!\langle j|, \qquad
    \hat{E}^+ = \sum_j |j+1\rangle\!\langle j|,
\end{equation}
with $\hat{E}^-=(\hat{E}^+)^\dagger$. Using properties of Bessel functions of the first kind, one finds
\begin{equation}\label{besself}
    \hat{D}(\beta)|j\rangle =
    \sum_{l=-\infty}^{+\infty}
    J_l(2|\beta|)\,e^{i l \phi}\,|j+l\rangle,
\end{equation}
where $\phi=\arg\beta$. Employing $J_{-l}(x)=(-1)^l J_l(x)$, the displaced Fock state is symmetric and localized around site $j$ with effective width $2|\beta|$. Thus, whereas $\hat{D}(\beta)$ acts as a displacement in the LPS, it modifies the width rather than shifting the center in the FSL.

For simplicity we take $\beta$ real. Owing to translational invariance, the root generators are diagonal in the momentum basis, $\hat{E}^\pm|k\rangle=e^{\pm i k}|k\rangle$, which allows diagonalization of the walker~(\ref{walkerun}) and full unitary~(\ref{fullun}). In momentum space,
\begin{equation}
    \hat{U}_w(k)
    =
    e^{i2\beta\sin k}\,|\uparrow\rangle\!\langle\uparrow|
    +
    e^{-i2\beta\sin k}\,|\downarrow\rangle\!\langle\downarrow|,
\end{equation}
so that
\begin{equation}
    \hat{W}(k)=\frac{1}{\sqrt{2}}
    \begin{pmatrix}
        e^{i2\beta\sin k} & e^{i2\beta\sin k} \\
        e^{-i2\beta\sin k} & -e^{-i2\beta\sin k}
    \end{pmatrix}.
\end{equation}
The eigenvalues $\lambda_\pm(k)=e^{\pm i\omega(k)}$ satisfy
\begin{equation}
    \cos\omega(k)
    =
    \frac{\cos\!\left(2\beta\sin k\right)}{\sqrt{2}}.
\end{equation}

With $|w_\pm(k)\rangle$ the eigenstates of $\hat{W}(k)$, we write
\begin{equation}
    \hat{W}^m(k)
    =
    \sum_{\pm}
    e^{\pm i m \omega(k)}
    |w_\pm(k)\rangle\!\langle w_\pm(k)|.
\end{equation}
For an initial state $|j_0\rangle\otimes|\chi_0\rangle$, where
$|j_0\rangle=\int_{-\pi}^{\pi}\frac{dk}{2\pi}e^{-ik j_0}|k\rangle$, the state
after $m$ steps becomes
\begin{equation}
\begin{aligned}
|\psi(m)\rangle
=
\int_{-\pi}^{\pi}\frac{dk}{2\pi}
\,e^{-ik j_0}
|k\rangle \otimes
\Big[
c_{+}(k)\, e^{i\omega(k)m}\,|w_{+}(k)\rangle \\
+
c_{-}(k)\, e^{-i\omega(k)m}\,|w_{-}(k)\rangle
\Big],
\end{aligned}
\end{equation}
with $c_\pm(k)=\langle w_\pm(k)|\chi_0\rangle$.

The walker probability distribution is therefore
\begin{equation}
P(j,m)
=
\Bigg|
\int_{-\pi}^{\pi}\frac{dk}{2\pi}
\Big[
c_{+}(k) e^{i\Phi_{j,m}(k)}
+
c_{-}(k) e^{-i\Phi_{j,m}(k)}
\Big]
\Bigg|^2,
\end{equation}
where $\Phi_{j,m}(k)=m\omega(k)-k(j-j_0)$. For large $m$, the dominant contributions arise from stationary points satisfying $\partial_k\Phi_{j,m}(k)=0$. Defining the group velocity $v_{\mathrm g}=d\omega/dk$, this condition implies
\begin{equation}
    v_{\mathrm g}=\frac{j-j_0}{m}.
\end{equation}
If $v_{\max}$ denotes the maximal group velocity, the distribution is restricted to $|j-j_0|\le v_{\max}m$, demonstrating ballistic spreading.

An additional remarkable feature is that the walker distribution is independent of the initial coin state. This can be seen directly from Eq.~(\ref{besself}). If $\psi_{\uparrow,\downarrow}(j,m)$ denotes the amplitude at site $j$ after $m$ steps for either coin component, the update
rule takes the convolution form
\begin{equation}
    \psi_{\uparrow,\downarrow}(j,m+1)
    =
    \sum_l A_{j-l}\,
    \psi_{\uparrow,\downarrow}(l,m).
\end{equation}
Hence,
\begin{equation}
    \psi_{\uparrow,\downarrow}(j,m)
    =
    \sum_l K_{j-l}(m)\,
    a_{\uparrow,\downarrow}(l),
\end{equation}
where $a_{\uparrow,\downarrow}(l)$ are the initial amplitudes. For an initial Fock state $|j_0\rangle$, $a_{\uparrow,\downarrow}(l)\propto\delta_{j_0,l}$. Because the kernel $K_{j-l}(m)$ is identical for both coin components, the walker distribution reduces to
\begin{equation}
    P(j,m)=|K_{j-j_0}(m)|^2,
\end{equation}
independent of the initial coin state. This does not imply factorization of coin and walker degrees of freedom; on the contrary, the coin--walker entanglement grows rapidly and depends sensitively on the initial coin state.

Another noncompact algebra that yields a distinctive DTQW on the FSL is $\mathfrak{su}(1,1)$. A version of the DTQW on the LPS generated by the Cartan operator was studied in~\cite{duan2023quantum}. However, since the Cartan generator is diagonal in the Fock basis, it does not induce a nontrivial walk on the FSL. We therefore construct the walker unitary from the root generators, as in the previous examples. Several bosonic representations of $\mathfrak{su}(1,1)$ exist; we employ the common realization~\cite{buvzek19901}
\begin{equation}\label{smsu11}
    \hat{K}_0 = \frac{1}{2}\left(\hat{a}^\dagger\hat{a} + \frac{1}{2}\right), \qquad
    \hat{K}^+ = \frac{1}{2}(\hat{a}^\dagger)^2, \qquad
    \hat{K}^- = \frac{1}{2}\hat{a}^2,
\end{equation}
which satisfies the commutation relations
\begin{equation}
    [\hat{K}^+,\hat{K}^-] = -2\hat{K}_0, \qquad
    [\hat{K}_0,\hat{K}^\pm] = \pm \hat{K}^\pm.
\end{equation}
Although the structure resembles that of $\mathfrak{hw}$, the minus sign in the first commutator fundamentally alters the geometry. The LPS is no longer flat but corresponds to a curved hyperboloid determined by the Casimir operator
$\hat{K}^2=\hat{K}_0^2-\hat{K}_1^2+\hat{K}_2^2$, where
$\hat{K}_1=(\hat{K}^++\hat{K}^-)/2$ and
$\hat{K}_2=i(\hat{K}^--\hat{K}^+)/2$. The Fock states are denoted $|k,m\rangle$, with Casimir eigenvalues $k=1/4$ or $k=3/4$. These correspond to two distinct LPSs depending on whether the vacuum $|0\rangle$ or the one-boson state $|1\rangle$ is chosen as reference.

This model illustrates a case in which the LPS differs conceptually from the $x$--$p$ phase space of a single bosonic mode. Acting with the displacement
operator~(\ref{gendis}) on the reference state $|0\rangle$ generates a squeezed vacuum state~\cite{mandel1995optical}. In the $x$--$p$ phase space the Husimi function of this state is a centered, anisotropic minimum-uncertainty
Gaussian, whereas on the LPS it corresponds to a localized Husimi distribution displaced a geodesic distance $d_{\mathfrak{su}(1,1)}$ from the origin, as given by Eq.~(\ref{psdist}). Thus the mapping between LPS and FSL representations is nonlinear.

For a squeezed vacuum with real squeezing parameter $\beta$, the FSL distribution is~\cite{mandel1995optical}
\begin{equation}
    P(2n) = \frac{(2n)!}{2^{2n}(n!)^2}\,
    \frac{\tanh^{2n}\beta}{\cosh\beta},
\end{equation}
so that only even Fock states are populated (reflecting the two reference parities of the algebra). The mean boson number and its variance are
\begin{equation}\label{squeezavg}
\begin{aligned}
    \langle\hat{n}\rangle &= \sinh^2\beta, \\
    \Delta n^2 &= 2\sinh^2\beta\,(\sinh^2\beta+1).
\end{aligned}
\end{equation}
The distribution therefore decays exponentially with $n$
($P(2n)\sim e^{2n\ln(\tanh\beta)}/\sqrt{2n}$) while its width grows with $\beta$; the vacuum state always carries the highest population. Consequently, a localized Perelomov coherent state on the LPS maps to a delocalized state on
the FSL, providing a clear example of a nonlinear correspondence between the two representations.

In the small-$\beta$ limit the Trotter approximation~(\ref{trotter}) applies, and the DTQW reduces to continuous evolution generated by a parametric amplification (squeezing) Hamiltonian~\cite{caves1982quantum,
collett1984squeezing}. The averages then follow Eq.~(\ref{squeezavg}), and the walker spreads with variance $\Delta n^2\sim e^{4m\beta}$, indicating super-ballistic expansion. Numerical simulations confirm this behavior; for
larger $\beta$ the spreading remains exponential but with a reduced effective exponent.


\begin{figure}[!ht] 
    \centering
    \includegraphics[width=0.95\linewidth]{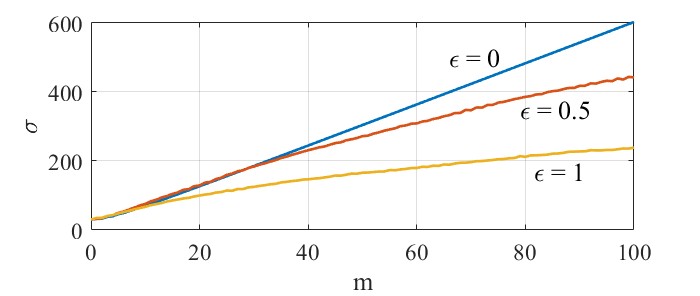} 
    \caption{Evolution of the width $\sigma=\sqrt{\langle\hat{n}^2\rangle-\langle\hat{n}\rangle^2}$ of the $\mathfrak{hw}$ walker for different coin dephasing strengths $\epsilon$. Dephasing is modeled by applying an additional rotation $\hat{R} (\epsilon)=\exp(i\epsilon\hat{\sigma}_z)$ at each coin toss, in addition to the Hadamard operation, where the angle $\epsilon$ is drawn independently at every step from a normal distribution with standard deviation $\epsilon$. The data shown are averages over 400 stochastic realizations. The walker is initialized in a coherent state with $\alpha=30$. The figure demonstrates the crossover from ballistic spreading for unitary evolution ($\epsilon=0$) to classical diffusive behavior for strong dephasing ($\epsilon=1$). }
    \label{fig:dwidth}
\end{figure}

\section{Discussion and concluding remarks}\label{sec:con}
Moving the idea of a DTQW from a discretized real-space lattice to state space is natural because Fock-state lattices (FSLs) typically lead to clean lattice models. The difficulty of implementing the walker unitary~(\ref{wshift}), which requires a highly nonlocal Hamiltonian, remains; we therefore employed displacement operators~(\ref{gendis}) instead. These operators do not shift the walker by exactly one site but generate finite amplitudes to tunnel to multiple sites.

Moreover, the displacement operator acts differently depending on the
location in the FSL, as is especially evident from the $\mathfrak{su}(2)$
example and Fig.~\ref{fig:ps_fsl}. This behavior originates from the broken
translational invariance of FSLs, which results from state-dependent tunneling amplitudes governed by matrix elements of the root generators. Our DTQW construction is purely algebraic, with the FSL sites identified with the
weight lattice. In physical lattice models, however, tunneling amplitudes are
often related to intersite distances $d_n$; in a first approximation one may
take $J_n \sim 1/d_n$. The FSL could then be rescaled so that distances are
defined by the magnitudes $|J_n|$. Caution is required: for rank-1 algebras the FSL is a simple tight-binding chain (one root pair), and this rescaling is straightforward, but for rank-2 algebras next-nearest-neighbor tunneling may prevent embedding the rescaled lattice in a plane. For example,
$\mathfrak{so}(5)$ produces a geometry that cannot be flattened, whereas the
triangular FSL of $\mathfrak{su}(3)$ corresponds to a curved plane. Such
rescaled FSLs therefore provide a natural setting for simulating DTQWs on
curved surfaces.

The main motivation for phase-space implementations of DTQWs is that
experimental realization of the walker unitary is comparatively
straightforward~\cite{xue2009quantum,schmitz2009quantum,
zahringer2010realization,sanders2003quantum,xue2008quantum,
flurin2017observing}. Likewise, DTQWs on FSLs can be interpreted in the LPS,
and displacement operators~(\ref{gendis}) are often easier to realize than the nonlocal Hamiltonian~(\ref{wham}). Most of the DTQWs considered here involve multiple bosonic modes, complicating experimental implementation. We briefly discuss possible realizations, noting that the $\mathfrak{hw}$ algebra corresponds directly to standard phase-space DTQWs.

The $\mathfrak{su}(2)$ algebra can be coupled to a two-level coin using bosonic modes in cavity or circuit QED or trapped-ion systems~\cite{larson2021jaynes}. Consider two degenerate photon modes, $\omega_a=\omega_b$, dispersively coupled to a two-level atom. After adiabatic elimination (and ignoring constant terms), the effective interaction Hamiltonian becomes
\begin{equation}
    \hat{H}_{2,\mathrm{eff}}
    = \chi\left(\hat{a}^\dagger\hat{b}+\hat{b}^\dagger\hat{a}\right)\hat{\sigma}_z,
\end{equation}
where $\chi$ is the two-photon coupling. This Hamiltonian describes virtual
photon scattering between the modes and has the desired structure for the
walker unitary. A two-photon generalization, replacing
$\hat{b}^\dagger \rightarrow \hat{a}$, yields the displacement operator for the $\mathfrak{su}(1,1)$ algebra.

Generalizing to $\mathfrak{su}(3)$ is less direct because a six-level coin is
required to couple selectively to three bosonic modes. One approach is to
exploit polarization and selection rules for the modes; an alternative is to
use three single-mode cavities arranged in a triangular configuration so that
neighboring cavities overlap~\cite{leonard2017monitoring}. Each node then hosts a two-level coin dispersively coupled to the two overlapping modes. Rather than a six-state coin, this setup employs three spatially separated two-state coins. The resulting coin Hilbert space is eight-dimensional (not six), so the Grover diffusion operator must be implemented with care. A similar strategy could be applied to the $\mathfrak{so}(5)$ model with four bosonic modes, either through spatial multiplexing or digital Trotterization.

Decoherence and experimental imperfections such as losses and dephasing are inevitable. As in conventional DTQWs, decoherence suppresses interference, driving a crossover from ballistic to diffusive spreading~\cite{brun2003quantum,kendon2007decoherence}. The same effect occurs for DTQWs on FSLs, as shown in Fig.~\ref{fig:dwidth} for the Heisenberg–Weyl case. Here we introduce stochastic coin dephasing at each step by applying a rotation $\hat{R}(\epsilon)=\exp(i\epsilon\hat{\sigma}_z)$ with $\epsilon$ drawn from a normal distribution of width $\epsilon$. Averages over 400 realizations demonstrate the expected transition from ballistic to diffusive broadening as the noise strength increases. Similar behavior arises when the walker itself is subject to dephasing. For quantum-optical implementations, boson loss is particularly important because it changes the dimensionality of the FSL. A more complete treatment employs the Liouville Fock-state lattice formalism introduced in~\cite{naves2025liouville}, which extends the FSL concept to density matrices and enables analysis of coherence decay.

Another feature emerging in the present framework is a form of localization that arises in certain multi-dimensional walks. In particular, for the $\mathfrak{su}(3)$ and $\mathfrak{so}(5)$ models we find that when the displacement amplitude $\beta$ becomes small while the effective time $T=m\beta$ is kept fixed, the walker dynamics becomes strongly suppressed and the distribution remains localized around the initial site. This behavior can be understood as a consequence of the rapid coin dynamics compared to the walker motion, which effectively averages the conditional shifts and cancels the leading-order contributions in the continuum limit. As a result, spreading occurs only through higher-order processes that scale as $\beta^2$, leading to an apparent freezing of the walker for small $\beta$. This mechanism is distinct from the well-known Grover localization in standard DTQWs, where only a fraction of the state becomes trapped due to flat bands in the dispersion relation.

Our study focused on representative Lie algebras; the list is far from exhaustive. Higher-rank algebras such as $\mathfrak{su}(n)$ and $\mathfrak{so}(n)$ generate higher-dimensional FSLs and richer DTQW dynamics. The noncompact $\mathfrak{su}(1,1)$ algebra illustrates an anomalous walk: Perelomov coherent states on the LPS correspond to squeezed states that are only weakly localized on the FSL. A related example is the symplectic algebra
$\mathfrak{sp}(2n,\mathbb{R})$, which describes two-mode squeezing with the
bosonic representation
\begin{equation}
\begin{array}{l}
    \hat{K}_0=\tfrac{1}{2}(\hat{n}_a+\hat{n}_b+1),\quad
    \hat{K}^+=\hat{a}^\dagger\hat{b}^\dagger,\quad
    \hat{K}^-=\hat{a}\hat{b},\\
    \hat{J}_z=\tfrac{1}{2}(\hat{n}_a-\hat{n}_b),\quad
    \hat{J}^+=\hat{a}^\dagger\hat{b},\quad
    \hat{J}^-=\hat{a}\hat{b}^\dagger .
\end{array}
\end{equation}
Because of a parity symmetry, the FSL decomposes into two square lattices. Combining these with the two-dimensional $\mathfrak{hw}$ algebra (adding $\hat{a}$, $\hat{a}^\dagger$, $\hat{b}$, and $\hat{b}^\dagger$) gives the \textit{Jacobi algebra}, where the parity symmetry is broken and now the FSL includes also next-nearest-neighbor tunneling, similar to $\mathfrak{so}(5)$ but in a noncompact setting. The displacement operators (\ref{gendis}) then incorporate both squeezing and displacement effects.

Lie superalgebras~\cite{scheunert2006theory} provide a natural
extension because many physical systems combine bosonic and fermionic degrees of freedom, for example in light–matter interacting models. Superalgebras incorporate both even and odd generators, combining commutation and anticommutation relations. Such structures could enrich DTQW dynamics on FSLs, but their detailed exploration lies beyond the scope of this work. 

In conclusion, we have introduced a framework for DTQWs on FSLs that generalizes standard DTQWs to state-space geometries defined by Lie algebras. The algebraic structure of the FSL determines the geometry of the walk and its interference properties. While experimental realization poses challenges, particularly for multi-mode systems and higher-rank algebras, the use of displacement operators and phase-space interpretations provides a more promising avenue for implementations. Future work may explore larger algebraic structures, nontrivial lattice embeddings, and the role of decoherence in extended Liouville FSLs, with potential applications to quantum simulation and quantum information processing.

\end{document}